\documentclass[]{spie}  

\usepackage{amsmath,amsfonts,amssymb}
\usepackage{graphicx}
\usepackage[colorlinks=true, allcolors=blue]{hyperref}

\newcommand{\strobex}{\textit{STROBE-X}}

\title{STROBE-X: A probe-class mission for X-ray spectroscopy and timing on timescales from microseconds to years}

\author[a]{Paul~S.~Ray}
\author[b]{Zaven~Arzoumanian}
\author[c]{S\o{}ren~Brandt}  
\author[b]{Eric~Burns}
\author[d]{Deepto~Chakrabarty}
\author[e,f]{Marco~Feroci}
\author[b]{Keith~C.~Gendreau}
\author[g]{Olivier~Gevin}
\author[h]{Margarita~Hernanz}
\author[i]{Peter~Jenke}
\author[b]{Steven~Kenyon}
\author[h]{Jos\'e~Luis~G\'alvez}
\author[j]{Thomas~J.~Maccarone}
\author[b]{Takashi~Okajima}
\author[d]{Ronald~A.~Remillard}
\author[g]{St\'ephane~Schanne}
\author[k]{Chris~Tenzer}
\author[l]{Andrea~Vacchi}
\author[m]{Colleen~A.~Wilson-Hodge}
\author[n]{Berend~Winter}
\author[n]{Silvia~Zane}

\author[o]{David~R.~Ballantyne}
\author[p]{Enrico~Bozzo}
\author[q]{Laura~W.~Brenneman}
\author[r]{Edward~Cackett}
\author[e]{Alessandra~De~Rosa}
\author[s]{Adam~Goldstein}
\author[t]{Dieter~H.~Hartmann}
\author[d]{Michael~McDonald}
\author[u]{Abigail~L.~Stevens}
\author[v]{John~A.~Tomsick}
\author[w]{Anna~L.~Watts}
\author[x]{Kent~S.~Wood}
\author[y]{Abderahmen~Zoghbi}
\author[z]{the STROBE-X Science Working Group}

\affil[a]{U.S. Naval Research Laboratory, Washington, DC, USA}
\affil[b]{NASA Goddard Space Flight Center, Greenbelt, MD, USA}
\affil[c]{National Space Institute, Technical University of Denmark, Denmark} 
\affil[d]{MIT Kavli Institute for Astrophysics and Space Research, Cambridge, MA, USA}
\affil[e]{IAPS/INAF, Rome, Italy}
\affil[f]{INFN/Roma Tor Vergata, Rome, Italy}
\affil[g]{CEA Paris-Saclay, DRF/IRFU, Gif sur Yvette, France}
\affil[h] {Institute of Space Sciences (ICE, CSIC) and IEEC, Bellaterra (Barcelona), Spain}
\affil[i]{University of Alabama in Huntsville, Huntsville, AL, USA}
\affil[j]{Department of Physics \& Astronomy, Texas Tech University, Lubbock TX, USA}
\affil[k]{IAAT, University of Tuebingen, Germany}
\affil[l]{Univ. of Udine \& INFN/Trieste, Italy}
\affil[m]{Astrophysics Branch, NASA Marshall Space Flight Center, Huntsville, AL, USA}
\affil[n]{Mullard Space Science Laboratory, University College London, Dorking, Surrey, UK}
\affil[o]{Center for Relativistic Astrophysics, Georgia Institute of Technology, Atlanta, GA, USA}
\affil[p]{Department of Astronomy, University of Geneva, Versoix, Switzerland}
\affil[q]{Smithsonian Astrophysical Observatory, Cambridge, MA, USA}
\affil[r]{Department of Physics \& Astronomy, Wayne State University, Detroit, MI, USA}
\affil[s]{Science and Technology Institute, USRA, Huntsville, AL, USA}
\affil[t]{Department of Physics \& Astronomy, Clemson University, Clemson, SC, USA}
\affil[u]{Department of Physics \& Astronomy, Michigan State University, East Lansing, MI, USA}
\affil[v]{Space Sciences Laboratory, University of California, Berkeley, CA, USA}
\affil[w]{Anton Pannekoek Institute for Astronomy, University of Amsterdam, The Netherlands}
\affil[x]{Praxis, Inc, Arlington, VA, USA}
\affil[y]{Department of Astronomy, University of Michigan, Ann Arbor, MI, USA}
\affil[z]{\url{https://gammaray.nsstc.nasa.gov/Strobe-X}}
\authorinfo{Send correspondence to P.S.R, E-mail: paul.ray@nrl.navy.mil}

\begin{document} 
\maketitle

\clearpage

\begin{abstract}
We describe the Spectroscopic Time-Resolving Observatory for Broadband
Energy X-rays ({\em STROBE-X}), a probe-class mission concept that will
provide an unprecedented view of the X-ray sky, performing timing and
spectroscopy over both a broad energy band (0.2--30 keV) and a wide
range of timescales from microseconds to years.
{\em STROBE-X} comprises two narrow-field instruments and a wide field
monitor. The soft or low-energy band (0.2--12 keV) is covered by an
array of lightweight optics (3-m focal length) that concentrate
incident photons onto small solid-state detectors with CCD-level
(85--175 eV) energy resolution, 100 ns time resolution, and low
background rates. This technology has been fully developed for {\em NICER}
and will be scaled up to take advantage of the
longer focal length of {\em STROBE-X}.  The higher-energy band (2--30 keV) is covered by large-area, collimated silicon drift
detectors that were developed for the European {\em LOFT} mission concept.
Each instrument will provide an order of magnitude improvement in
effective area over its predecessor ({\em NICER} in the soft band and {\em RXTE}
in the hard band). Finally, {\em STROBE-X} offers a sensitive wide-field
monitor (WFM), both to act as a trigger for pointed observations of
X-ray transients and also to provide high duty-cycle, high time-resolution, 
and high spectral-resolution monitoring of the variable X-ray
sky.  The WFM will boast approximately 20 times the sensitivity of the {\em RXTE} All-Sky Monitor,
enabling multi-wavelength and multi-messenger investigations with a large
instantaneous field of view. This mission concept will be presented to the 2020 Decadal Survey for consideration.
\end{abstract}


\section{INTRODUCTION}
\label{sec:intro}  

High-throughput timing and spectroscopy in the X-ray band is an extremely powerful technique for probing the physics 
and astrophysics of energetic sources throughout the Universe, ranging from nearby stars and compact objects in our Galaxy
to supermassive black holes and cosmic explosions at high redshift. The power of this technique was amply demonstrated by the highly successful
\textit{Rossi X-ray Timing Explorer} (\textit{RXTE}) mission, which operated from 1995 to 2012. {\em RXTE} employed gas proportional 
counter detectors and covered the X-ray band above 2 keV. Since then, the development of large area silicon drift detectors (SDDs) and
micropore X-ray collimators has enabled large area instruments that are much smaller and lighter per unit area. 
This technology was developed extensively for the European \textit{Large Observatory for X-ray Timing} ({\em LOFT}) mission during a 
multi-year assessment phase study as part the ESA M3 process. Simultaneously in the U.S., lightweight X-ray concentrators
paired with small SDDs were developed and deployed on the {\em NICER} mission, demonstrating that similar advances were possible
in the soft band (0.2--12 keV, with effective area peaking around 1.5 keV). 

Here we present the 
\textit{Spectroscopic Time-Resolving Observatory for Broadband Energy X-rays} (\strobex{}), which brings these two technologies 
together into a uniquely powerful observatory that has over an order of magnitude more effective area than {\em NICER} in the soft band and
{\em RXTE} in the hard band. This combination allows simultaneous measurements of thermal and non-thermal emission processes and precise
characterization of the relationship between the two, and it also enables the effects of absorption to be clearly separated from 
continuum emission. A third instrument monitors the X-ray sky over a range of timescales that will both trigger
observations by the pointed instruments and be a powerful instrument for discovering and characterizing rare transients on its own. In the following sections, we describe the science drivers, the designs of each of the three instruments, and the mission design resulting from our studies in the NASA/GSFC Integrated Design Center (IDC).

\begin{figure}
\centering\includegraphics[width=4.0in]{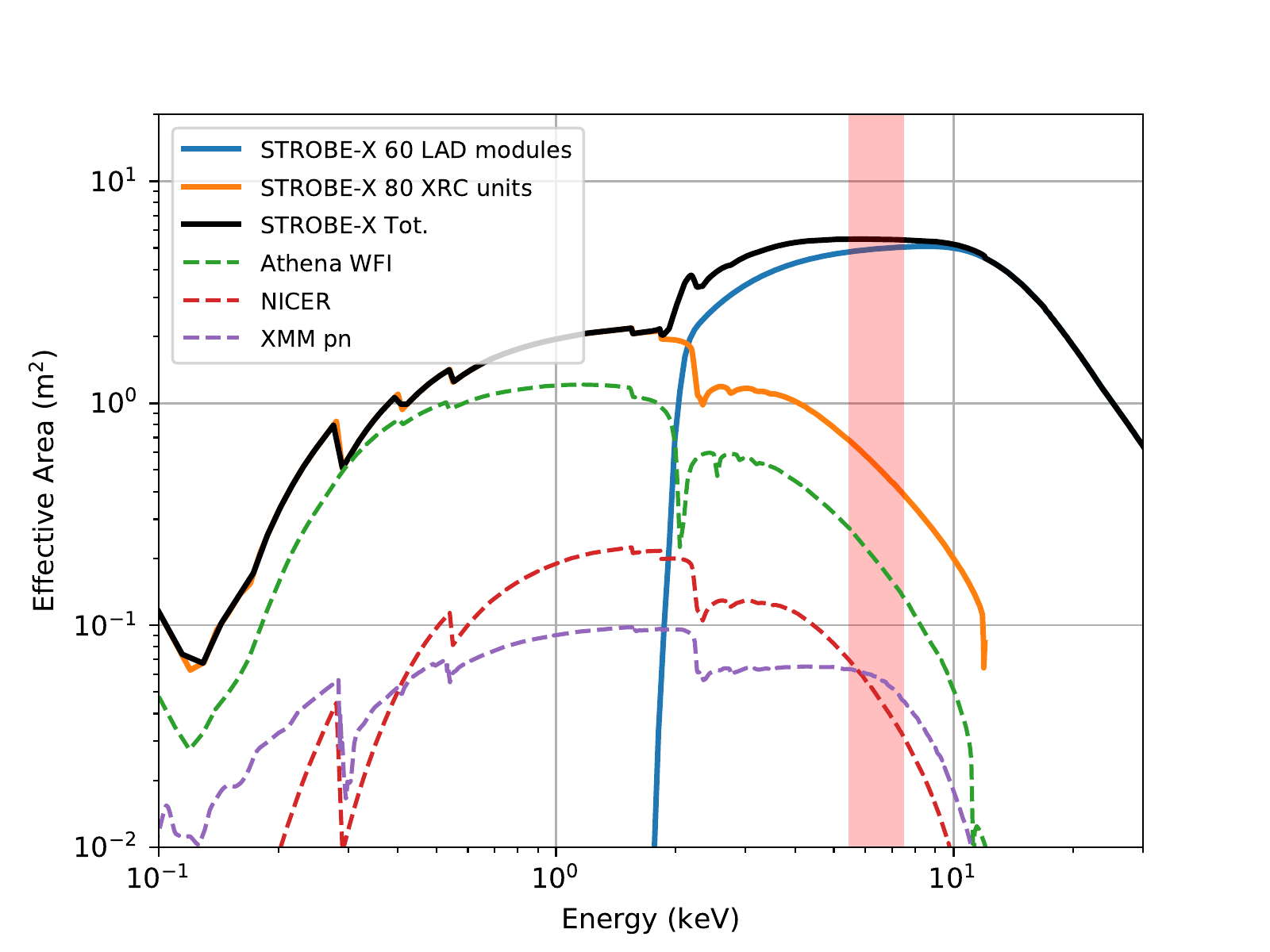}
\caption{Effective area of the {\em STROBE-X} pointed instruments (solid curves), compared to some previous and planned missions (dashed curves). {\em STROBE-X} has the largest area over its entire bandpass. The Fe-K line region near 6.4 keV is denoted by the pink band. 	\label{fig:effarea}}
\end{figure}

\section{SCIENCE DRIVERS}

The {\em STROBE-X} mission includes a versatile set of instruments optimized for fast timing and broadband, time-resolved spectroscopy of compact objects.  The key science drivers are threefold: (1) measuring the spin distribution of accreting black holes, (2) understanding the equation of state of dense matter, and (3) exploring the properties of the precursors and electromagnetic counterparts of gravitational wave sources.  A wide range of additional science is also enabled by {\em STROBE-X's} unique combination of instruments, from studies of the inner solar system to the high redshift Universe.  Herein we outline the key science cases, with particular attention given to probing the counterparts of gravitational wave sources, as this area of study has expanded rapidly in the past two years.

\subsection{Key goals}
\subsubsection{Black Hole Spins}

There are strong astrophysical motivations for measuring black hole (BH) spins.  A slowly-rotating accreting BH can only reach the maximal spin allowed in general relativity by doubling its mass\cite{Bardeen_1970}.  Understanding the spin distribution of stellar-mass BHs (where such growth is unlikely) thus yields essential information on the BH formation process, and hence on the evolution of massive stars and the supernova mechanism\cite{Miller_etal_2011,Fragos_McClintock_2015}.  A comparison of the birth spin rates to the distribution of BH spins at the ends of their binary phase constrained via gravitational wave measurements\cite{Abbott_etal_2016,Farr_etal_2017} is of special interest.  In the case of supermassive BHs in active galactic nuclei (AGN), the distribution of spins encodes many epochs of growth via gas accretion and mergers with other BHs, thereby offering an intriguing probe of galaxy formation and evolution (see, e.g., Ref.~\citenum{Berti_Volonteri_2008}).

{\em STROBE-X} is designed to probe the spins of BHs in several different ways.  At the present time, BH spins are usually estimated either through spectral fitting of the accretion disk thermal continuum (e.g., Refs.~\citenum{Remillard_McClintock_2006,Shafee_etal_2006,Steiner_etal_2009,McClintock_etal_2011}) or by modeling the relativistic smearing of the reflection spectrum of the accretion disk (e.g., Refs.~\citenum{Brenneman_Reynolds_2006,Reynolds_2014,Garcia_etal_2015}).  These methods can be applied using integrated spectra with modest count rates, though both methods require that the BH is actively accreting, either from the surrounding gas (in an AGN) or from a binary companion (in an X-ray binary, or XRB).

Although current spectroscopic and timing measurements have resulted in tens of published spin values for stellar-mass BHs in XRBs and their supermassive counterparts in AGN over the past decade, systematic uncertainties related to both the instruments and models used hamper our ability to draw conclusions from this sample (e.g., Refs.~\citenum{Brenneman_2013,Reynolds_2014,Choudhury_etal_2017,Taylor_Reynolds_2018}). {\em STROBE-X's} ability to simultaneously use different diagnostics to measure BH spin will allow us to overcome systematic uncertainties linked to spectral modeling in AGN studies.

\begin{figure}
\includegraphics[width=2.75in]{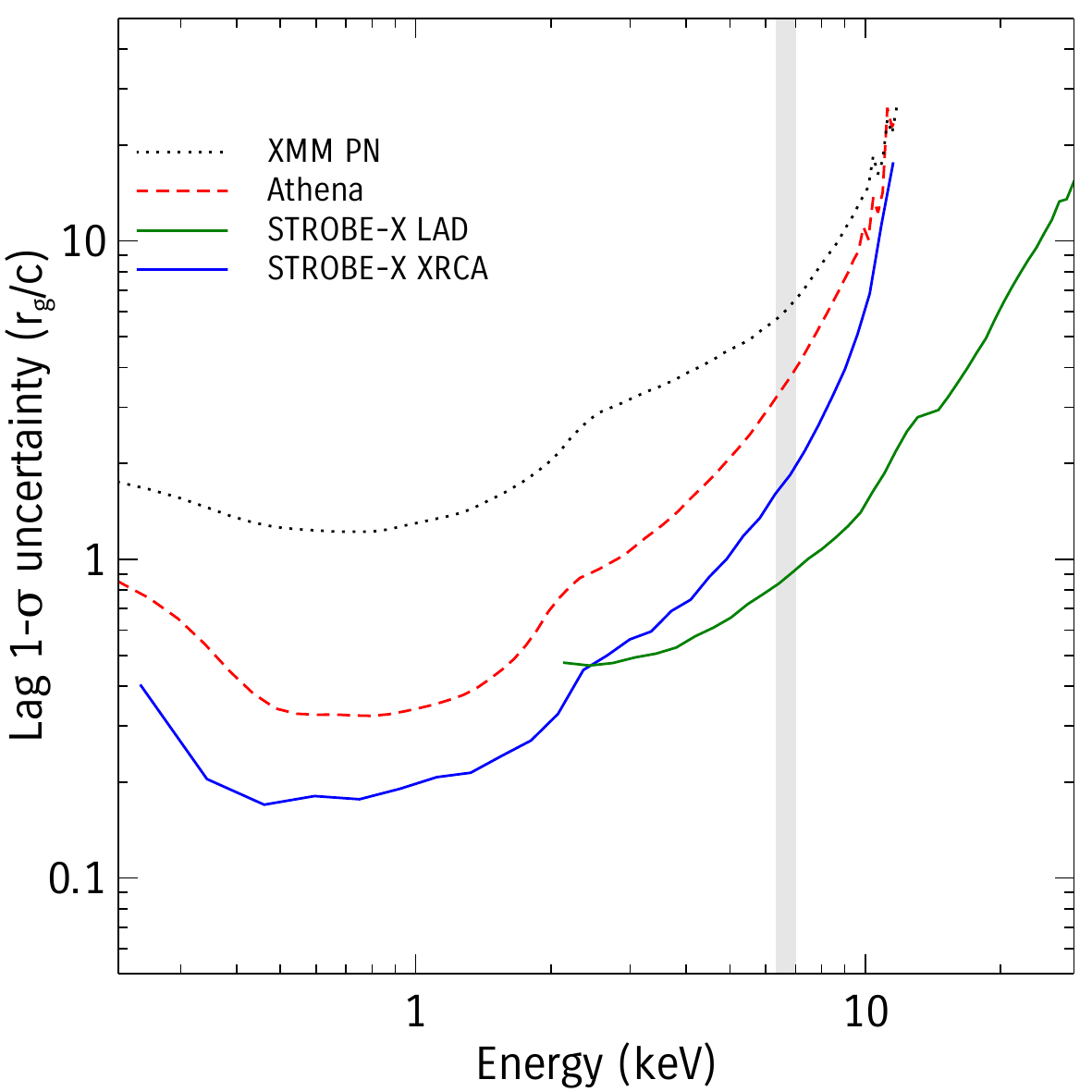}\hfill\includegraphics[width=3.5in]{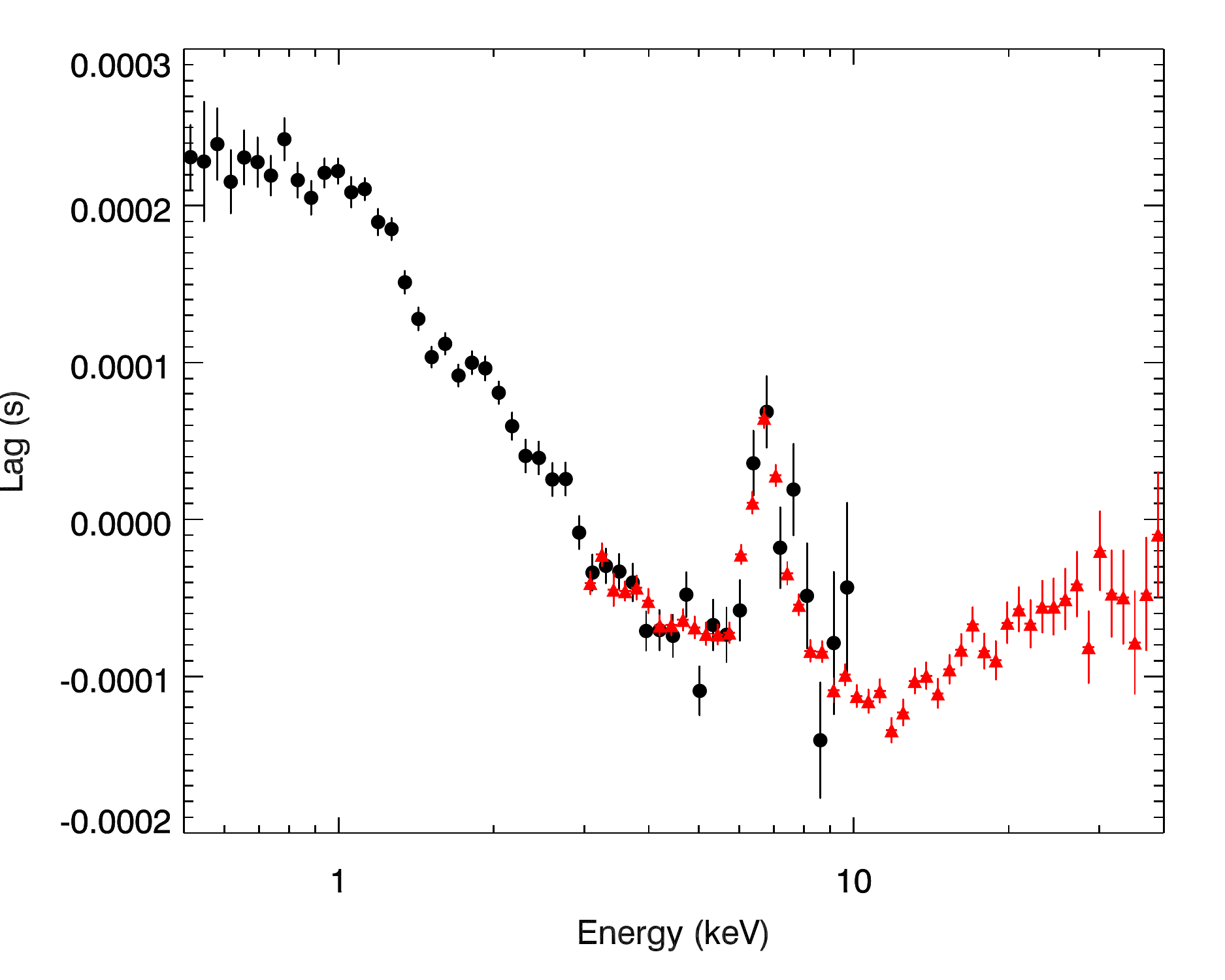}
\caption{Sensitivity of {\em STROBE-X} for reverberation mapping.  Left: {\em STROBE-X} (solid curves) compared with other satellites (dotted curves) for an AGN black hole, assuming a 2 mCrab source, 100 ks exposure, and a $10^6 M_\odot$ black hole. {\em STROBE-X} is more sensitive over the entire bandpass. The Fe-K line region near 6.4 keV is denoted by the grey band. Right:  A simulated 1 ks {\em STROBE-X} observation of the XRB black hole GX~339-4 showing 1--10 Hz reverberation lags (XRCA shown as black circles, LAD as red triangles), with parameters following those of Ref.~\citenum{Uttley_etal_2011}.  The LAD lag precision at the Fe-K line is about 20~$\mu$s per spectral bin in a 1~ks observation, significantly better than 1~$R_g/c$ (a gravitational radius crossing time) for a 10\,$M_\odot$ black hole. \label{fig:zoghbi_cackett_fig}}
\end{figure}

In the case of XRBs with higher count rates and more rapid variability, spin measurements will be possible with seconds, rather than hours of data, making it possible to determine whether the model parameters are stable as sources vary.  The use of relativistic reverberation mapping, which currently is strongly count-rate limited for XRBs, will provide an additional check on the reflection models by verifying that the time delays of different emission components compared to the continuum emission originate in different regions (namely the corona and disk), as expected for the combination of estimated spin and mass\cite{Uttley_etal_2014}. Figure~\ref{fig:zoghbi_cackett_fig} illustrates the capabilities of {\em STROBE-X} for measuring reverberation lags.

High frequency quasi-periodic oscillations (HFQPOs) will provide a  complementary approach to estimating BH spins.  The observed HFQPOs are often found to exist in 3:2 frequency ratios.  Under a given assumption about the mechanism for producing the QPOs, they can be used to place tight constraints on the BH spins (e.g., Ref.~\citenum{Motta_etal_2014}), but at the present time the results from HFQPO modeling, thermal continuum fitting and the reflection fitting method are not in universal agreement with one another, so it is vital to collect new and better data to understand the systematics in the methods.  It is also important to note that reliable measurements of supermassive BH spins have so far only been achieved with the reflection fitting method, whereas thermal continuum fitting, HFQPO modeling and reflection fitting have all yielded spin constraints in stellar-mass BHs.  Cross-calibrating these techniques on XRBs will therefore prove broadly beneficial, yielding greater confidence that the measured supermassive BH spins are accurate.

\subsubsection{Neutron Star Equation of State}

Densities in the cores of neutron stars can reach up to ten times that of normal nuclear matter. In addition to nucleonic matter in conditions of extreme neutron richness, neutron stars may contain stable states of strange matter, either bound in the form of hyperons or in the form of deconfined quarks. Neutron stars are unique laboratories for the study of strong and weak force physics in cold, ultradense matter (see Ref.~\citenum{Oertel17} for a recent review).

\begin{figure}
\centering
\includegraphics[width=3.5in]{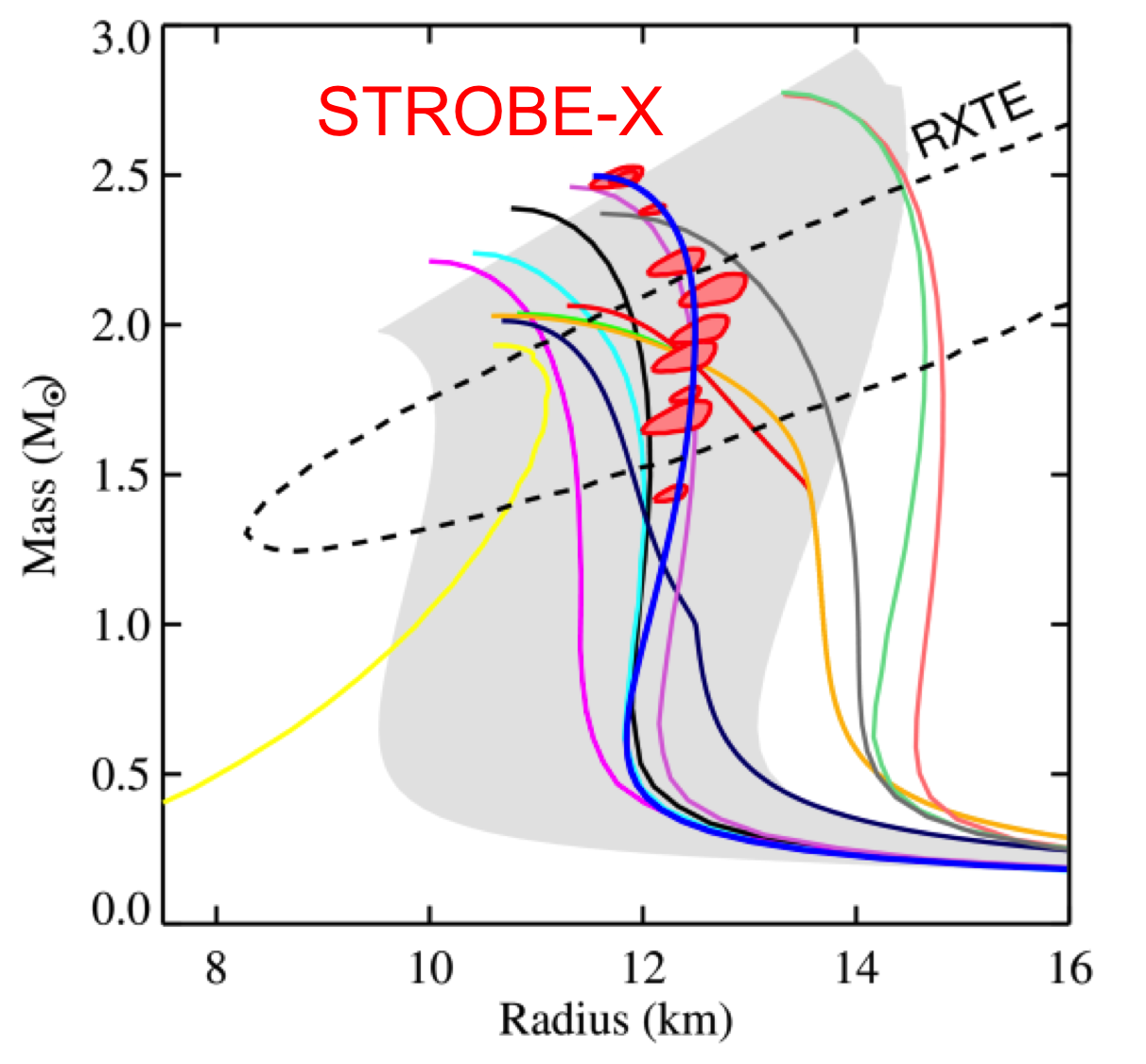}
\caption{Simulated neutron star mass-radius constraints obtained with {\em STROBE-X}. The red ellipses illustrate how 5\% measurements of $M$ and $R$ from many neutron stars, as expected from {\em STROBE-X}, will map out the full $M$-$R$ relation and thus tightly constrain the ultradense matter equation of state (EOS). The current {\em NICER} mission will only measure $\sim$4 neutron stars. An earlier {\em RXTE} constraint is also shown. The $M$-$R$ curve for the "true" EOS (blue in this example) must be consistent with all observations. The other colored curves are $M$-$R$ relations for other representative EOS models\cite{Lattimer_Prakash_2001}, and the grey band shows the range of EOS models based on chiral effective field theory\cite{Bednarek_etal_2012}.}	\label{fig:nseos}
\end{figure}

{\em STROBE-X} will deploy several techniques to study the equation of state (EOS) of ultradense matter in neutron stars.  The pulse profile modelling that can be done for a few pulsars with {\em NICER} (sufficient to provide a proof of concept\cite{Bogdanov13}) will be feasible for around 20 pulsars with {\em STROBE-X} (see Figure~\ref{fig:nseos}).   By sampling more neutron stars, we can study the EOS across a wider range of central densities, mapping the EOS more fully and probing with finer resolution any potential phase transitions. {\em STROBE-X} can achieve this because its large collecting area  enables us to apply the pulse profile technique to burst oscillation sources, which require long observation times to gather sufficient pulsed photons.  Burst oscillation sources are particularly useful: they have the most rapid spin rates of all the target sources (so that the relativistic effects exploited are larger); and have a well-understood beaming function\cite{Watts16}.  The larger area also enables us to target pulsations from accretion-powered pulsars, and thermally-emitting pulsars that are too faint for {\em NICER}.   The ability to target different source classes is very important, since it allows us to cross-check techniques (several stars have two different types of pulsations), compare different source populations, and hence identify and combat any model systematics.  This is a huge advance on {\em NICER}, which targets a single source class with a single type of pulsation.


The large area of {\em STROBE-X} will also deliver unprecedented sensitivity for measuring neutron star spin via the detection of accretion-powered pulsations and burst oscillations.  EOS models are associated with a certain maximum spin rate (break-up), and the discovery of a neutron star with a spin rate of $\sim$ 1 kHz or above would provide a very clean constraint on the EOS (the only model dependency being the assumption that GR is correct).  Transient, short-lived and weak pulsations could all be targeted.  At present the sample of accreting neutron stars with measured spins is small, and we have found pulsations all the way down to the sensitivity limit (see Ref.~\citenum{Watts16}).  

\subsubsection{Gravitational Wave Source Progenitors and Counterparts}

{\em STROBE-X} will provide a vital electromagnetic (EM) complement to all classes of gravitational wave studies.  In the high-frequency (LIGO/VIRGO third generation) bands, the WFM on {\em STROBE-X} would instantly provide positions accurate enough for ground-based spectroscopy with integral field units for the subset of sources which are either beamed toward Earth or within about 40 Mpc.  With the ground-based gravitational-wave community discussing trade-offs on the localization capability to deepen the detection horizon for third-generation instruments (e.g., Einstein Telescope, Cosmic Explorer), the prompt and precise localization of X-ray transient counterparts to compact binary mergers will be crucial.  Current estimates indicate that the binary neutron star detection rate by LIGO/Virgo will be $\lesssim$ 10/year for the next few years, and because some fraction of those may not have prompt EM signals, uncertainties related to the emission physics of short duration gamma-ray bursts will likely remain.  Indeed, the detection of GW170817 has raised the question about the nature of the EM emission---an off-axis structured relativistic jet or a trans-relativistic shock breakout---and there remains a considerable amount of uncertainty about the emission mechanism, jet physics, and range of properties of such mergers.  As shown in Figure~\ref{fig:grb_detections}, {\em STROBE-X} will have the capability to detect $\sim 10$ short duration gamma-ray bursts (compact binary mergers) per year, contributing to a larger statistical sample that can answer these questions.  

Another outstanding question is whether binary neutron star mergers immediately coalesce into a black hole or if there is a short-lived compact object, such as a magnetar, that exists before such a collapse.  For nearby mergers, {\em STROBE-X} will be able to detect such a short-lived object if it radiates soft X-rays or, alternatively, place constraining upper limits on such radiation. In addition to detecting binary neutron star mergers, gravitational-wave detectors are searching for merging neutron star-black hole systems, which might also be progenitors to some gamma-ray bursts.  The expected detection rate is $\lesssim O(1)$/year with current technology, though there are some expected sensitivity improvements to high-mass-ratio systems in third-generation detectors, so this will be a promising discovery space for {\em STROBE-X}.

For continuous wave sources (i.e., neutron stars with deformations leading to quadrupole moments -- e.g., Ref.~\citenum{Owen_etal_1998}), the LIGO positional precision for detections should be $\sim$10~arcsec. The strongest candidates for continuous gravitational wave sources should be fast-spinning young pulsars which tend to have the highest X-ray luminosities, or active accretors; the {\em STROBE-X}/XRCA is uniquely suited to detecting X-ray periodicities from these objects.

Looking forward, we can anticipate some overlap in time between {\em LISA} and {\em STROBE-X}.  Under conditions that will be met a small, but nontrivial fraction of the time\cite{2008ApJ...677.1184L}, {\em LISA} will obtain positions accurate to 10 square degrees and with redshifts accurate to a few percent $\sim$1 month before the mergers of supermassive black holes.  This will, in turn, allow searching for the counterparts with the Wide Field Monitor, and then after detection, with quick follow-up with XRCA.

The {\em STROBE-X}/WFM will be a unique instrument for finding merging supermassive black holes in a higher mass range than the one over which {\em LISA} is sensitive.  The detected sources are expected to be highly obscured, meaning that hard X-rays and radio are the best bands in which to study them\cite{2018MNRAS.tmp.1214B}.  Strong quasi-periodic oscillations, along with near-Eddington accretion, are also expected from double black holes as they merge\cite{2018MNRAS.476.2249T}\cite{2018ApJ...853L..17B}, which would be detectable out to redshift $z\sim$0.6 (where the mergers may be detectable in gravitational waves after the X-ray emission via their ringdowns\cite{2005MNRAS.361.1145R}; with the WFM, and, if one knew where to look, to redshift $z\sim 10$ with the XRCA.

\subsubsection{Gamma-ray Bursts and Supernovae}
\begin{figure}
\centering
\includegraphics[width=3.5in]{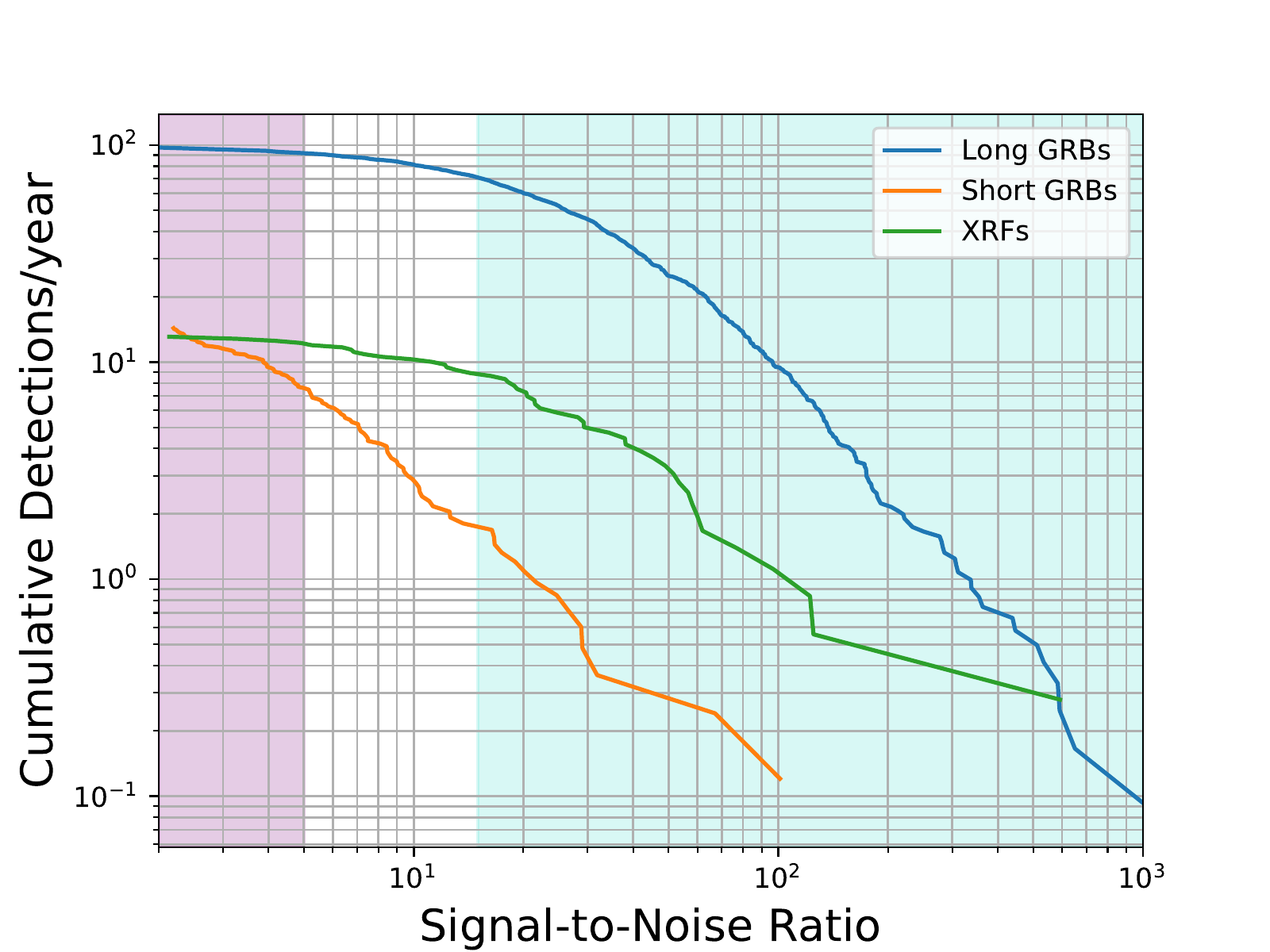}
\caption{The expected cumulative onboard detection rate of canonical gamma-ray bursts (GRBs) and X-ray flashes (XRFs) by the {\em STROBE-X}/WFM. It will detect $\sim 100$ long duration GRBs, $\sim 7$ short duration GRBs, and $\sim 12$ XRFs onboard per year. The onboard detection rate of long GRBs exceeds that of the {\em Swift}/BAT, while the short GRB detection rate is comparable.  A unique capability is the downlink of event data to the ground for the WFM, enabling sub-threshold searches to double the number of short GRB detections (purple shading).  The detection rate of XRFs exceeds that of previous instruments and is a particular science focus for the WFM.  The blue shading shows the region of signal-to-noise where high-fidelity spectroscopy can be performed in the prompt X-ray for these sources.\label{fig:grb_detections}}
\end{figure}

In addition to gravitational wave counterpart searches, {\em STROBE-X} will also dramatically enhance more traditional studies of gamma-ray bursts (GRBs) and supernovae, matching and exceeding the capabilities of the {\em Swift}/BAT.  The {\em STROBE-X}/WFM is similar to the {\em BeppoSax} Wide Field Camera but with a much larger field of view.  The population of relatively faint and soft GRBs called X-ray flashes (XRFs) has not been studied since the {\em BeppoSax} mission ended in 2002, due to the lack of wide-field, medium-energy X-ray instruments.  XRFs are still not well-understood, and may be normal GRBs viewed off axis or at high redshift, or they could be baryon-loaded ``dirty fireballs''\cite{1999ApJ...513..656D}.  As shown in Figure~\ref{fig:grb_detections}, the WFM is expected to detect more than 10 XRFs per year, most of them bright enough to perform detailed spectroscopy, which may help illuminate their relationship to canonical GRBs.  Additionally, the arcminute localization of the WFM will allow sufficiently precise positions to be sent immediately to the ground, facilitating optical/infrared spectroscopic follow-up using integral field units, which are expected to be commonplace on large ground-based telescopes by the 2030s. The automated repointing capability of {\em STROBE-X} will allow XRCA spectra for the late prompt emission for fortuitously located GRBs, and the early afterglows for most of them. 

Given its soft energy band, in principle the WFM detection efficiency for short/hard GRBs is expected to be rather low, with a detection rate of about 10 events/year (Figure~\ref{fig:grb_detections}). However, based on HETE-II and Swift/BAT measurements, there is evidence that at least a fraction of short GRBs show a weak soft extended emission following the first hard spike. The unprecedented combination of soft energy band, wide field of view and sensitivity of the WFM compared to previous and present GRB monitors, will allow it to detect tens of short GRBs per year through their soft extended emission, thus providing an additional channel for the detection and localization of events produced by NS-NS and NS-BH mergers.

Furthermore, X-ray spectroscopy of GRBs and early afterglows in the energy range where abundant metals have atomic lines and edges can help in a variety of ways.  Redshifts may be potentially obtained with prompt emission; this was done previously with {\em BeppoSAX}\cite{2000Sci...290..953A} but it could be done more frequently, precisely, and reliably with the higher spectral resolution of the WFM. Furthermore, searches for emission and absorption lines with the XRCA in the early afterglows could yield a host of information about the nature and environment of GRBs\cite{2002ApJ...572L..57L}.  The latter case has so far yielded only statistically marginal results, but could be pursued dramatically better with XRCA than with past instruments.

For supernovae, {\em STROBE-X} has the potential to be revolutionary.  Supernova shock breakouts should be visible to a distance of about 20 Mpc, and a few of these events are expected per year if the 2008 event in NGC~2770\cite{2008Natur.453..469S} is taken as a template.  The hard X-ray emission is the hottest and earliest signature of shock breakout.  Recently, an X-ray transient associated with a type Ic supernova at 60 Mpc would have been bright enough for several days for the WFM to detect, and may be a first example of an X-ray orphan afterglow\cite{2018ATel11737....1R}.  {\em STROBE-X} will discover similar events earlier relative to other proposed instruments (e.g., optical missions) that can also discover these shock breakouts, and its large instantaneous field of view will allow {\em STROBE-X} to discover the {\it nearest} of these events, offering unprecedented diagnostic tools. For example, shock breakout arrival times can be compared with the arrival times of neutrinos and gravitational waves to reveal the passage of blast waves through the envelopes of massive stars in their final moments. A full understanding of core-collapse supernovae has not yet been reached, and full multi-messenger data sets of these events in the local Universe will generate breakthroughs at this frontier. 

The advances made in the arenas of gravitational wave and neutrino detection have broadened our portfolio from multi-wavelengths to multi-messenger, and  time-domain astronomy is poised to create a new golden era of astrophysics. Probing early cosmic star formation with high-redshift GRBs, and exploring the dynamic physics of supernovae in the local Universe will be the focus of instruments that can respond to transients, probe their emission on the shortest timescales directly and provide precise locations and fast characterization to ground and space based observatories ready for follow-up. As demonstrated with missions such as {\em Swift} and {\em Fermi}, wide field-of-view and fast response are key elements of missions exploiting the era of time domain astronomy. {\em STROBE-X} offers unique capabilities that combines all-sky monitoring, sensitivity, rapid slewing and high time resolution, which by themselves and in combination with synergies with other observatories, will greatly advance our knowledge of transient phenomena such as GRBs, supernovae, shock breakouts, tidal disruption events, and others --- as we briefly discuss in the next section. 

\subsection{Ancillary science capabilities}

In addition to the core science goals envisioned for {\em STROBE-X}, the mission has unique capabilities to contribute to a wide variety of other scientific fields.

\subsubsection{An All-Sky Medium-Energy X-ray Survey}

{\em STROBE-X} would also complete an all-sky survey in the medium-energy X-rays that would surpass the current standard, the {\em RXTE} Slew Survey.  The {\em STROBE-X} survey would also fill in the gaps in the {\em RXTE} survey's coverage and, importantly, would improve the spectral resolution of that survey by a factor of $\sim5$.  In addition to providing sensitivity to highly obscured Galactic sources of moderate brightness, this survey would exceed the sensitivity of {\em eROSITA} to Compton-thick AGN by taking advantage of {\em STROBE-X's} superior collecting area, identifying the Compton-thick AGN using their very high equivalent width iron emission lines.  Furthermore, it would allow the discovery of the highest power, highest redshift blazars, as these objects would be detected by the {\em STROBE-X}/WFM but not by {\em Swift}/BAT in the same way that $z\sim4$ blazars are detectable by {\em Swift}/BAT but not by {\em Fermi}/LAT \cite{2011MNRAS.411..901G}.

\subsubsection{Accretion Flows and Accretion-Ejection Physics}

{\em STROBE-X} will probe the nature of accretion and how relativistic jets are ejected from actively accreting systems.  We will achieve a greater level of detail in detected quasi-periodic oscillations, and our studies of the ``noise'' components of power spectra of XRBs can help us understand both power spectra and QPOs and use them to probe accretion geometry.  Supermassive black hole mass estimates made from AGN power spectra\cite{2006Natur.444..730M} will be made at a variety of inclination angles, and can hence be used to tie together the masses from masers and those from reverberation mapping.

In addition, the broadband spectroscopic capabilities of {\em STROBE-X} will elucidate many aspects of AGN accretion disks, such as the origin of the so-called `soft excess'. This emission consists of an excess above the main power-law at energies $<$~1 keV, and the origin of the radiation remains mysterious, even three decades after its discovery. Two compelling models for the excess in unobscured AGNs are a `warm' corona \cite{2013A&A...549A..73P} that may be between the accretion disk and the hot corona producing the X-ray power-law, and highly blurred relativistic reflection from a high density disk that naturally produces a strong soft excess\cite{2004MNRAS.351...57B,2016MNRAS.462..751G}. The spectra predicted by these two models are almost identical between 1 and 10~keV, but the broadband spectra and high sensitivity of {\em STROBE-X} will easily be able to distinguish the two models in $\approx 20$~ks for a bright AGN. The short exposure time needed to make these measurements means that {\em STROBE-X} can quickly measure soft-excesses and, even more importantly, their variability in multiple AGN.  Moreover, the broad bandpass produced by the combination of data from the XRCA and LAD will yield measurements of the high-energy cutoff and subsequent hot coronal parameters (such as optical depth and electron temperature) in a fraction of the exposure time currently needed by {\em NuSTAR}.  This opens up the unique possibility of examining relationships between, e.g.,  the `hot' and `warm' coronas in many AGN.  Such an investigation could only be performed by {\em STROBE-X} and will be crucial to understanding the flow of accretion energy through the disk and the putative two-phase corona into the observed X-ray spectrum.

The fast broadband spectroscopy provided by {\em STROBE-X} will open new avenues for studying accretion physics far beyond the capabilities of prior missions.  For example, Type I X-ray bursts are nuclear explosions on the surfaces of accreting neutron stars that last less than a minute, but are expected to strongly interact with the surrounding accretion disk\cite{2005ApJ...626..364B,2018SSRv..214...15D}.  {\em STROBE-X} time-resolved spectroscopy of a Type I X-ray burst will be able to use X-ray reflection signatures to map out the response of the accretion disk to the burst in real time\cite{2016ApJ...826...79K}. As these bursts occur several times a day all over the sky, {\em STROBE-X} will be able to provide a never-before-seen view of the dynamics of accretion disks.

{\em STROBE-X} data collected in conjunction with data from other facilities can be used for multi-wavelength timing which can help illuminate the details of the disk-jet connection\cite{2015arXiv150102770D} \cite{2015arXiv150102766C}.  Jetted tidal disruption events (TDEs) can be discovered with {\em STROBE-X}\cite{2015arXiv150102774R}, and the details of the accretion flows in TDEs can be accurately characterized.  

\subsubsection{Abundance Measurements from High-Throughput Spectroscopy}

For a variety of classes of extended objects, the XRCA will provide  spectra of unprecedented signal-to-noise.  These include supernova remnants in the Magellanic Clouds (to allow classification of the type of supernova based on remnant abundances), comets and planets (to understand their composition and the composition of the solar wind as they interact through solar wind charge exchange and scattering), and low-redshift compact groups of galaxies (to understand chemical evolution in these environments).

\subsubsection{Stellar evolution}

{\em STROBE-X} will make key contributions to our understanding of the highest and lowest mass stars.  At the high-mass end, {\em STROBE-X} will probe the stellar remnant mass distribution, addressing the question of whether there is a gap between the most massive neutron stars and the least massive black holes\cite{Ozel10}\cite{Farr11} by dramatically growing the sample of XRBs with the WFM.  This instrument has greater sensitivity than previous wide-field X-ray monitors, paired with a hard enough response to be sensitive to highly obscured objects in the Galactic plane.  {\em STROBE-X} can also be used to understand the spin evolution of neutron stars as they accrete\cite{2015arXiv150102769M}.  The XRCA has the potential to detect pulsations from white dwarfs during the supersoft phases of their nova outbursts, potentially helping to reveal whether white dwarfs in CVs and symbiotic systems rotate fast enough to account for some of the spread in the observed nickel masses of Type Ia supernovae.

At the low-mass end of the spectrum, {\em STROBE-X} will provide a unique sensitivity to extreme stellar coronal flaring, with detection via the WFM\cite{2015arXiv150102771D}.  Characterization of the spectra of the flares using the fast slewing capability will revolutionize our understanding of extreme stellar flares.   

\subsubsection{Fundamental Physics}

{\em STROBE-X} is also poised to contribute to our knowledge of fundamental physics in multiple ways.  Its observations of long-duration nuclear bursts, especially given its capability to obtain XRCA and LAD data on the tail ends of these bursts, will provide empirical constraints on the properties of nuclear reactions of proton-rich isotopes, which is one of the central unanswered questions in nuclear physics \cite{2017ApJ...844..139S}.  {\em STROBE-X} will also bring a unique perspective to the search for dark matter, providing one of the few viable means of discovering axions through their interactions with strong magnetic fields \cite{2006PhRvD..74l3003L}.

\section{INSTRUMENTS}

The {\em STROBE-X} instrument suite has a strong heritage from the {\em NICER} mission in the U.S. and the {\em LOFT} mission concept that 
has been under study for many years in Europe. The team has created detailed designs and prepared thorough cost estimates
during a study at the NASA/GSFC Instrument Design Lab (IDL) in 2017 November and December. A major result of this study was the
division of the primary instrument into four identical ``quadrants,'' each with a composite optical bench for the XRCA and a 
deployable panel for the LAD. This design has several advantages: Firstly, integration and test flow is simplified, can incorporate 
parallelism, is reduced in cost, and 
requires smaller facilities than if the instruments were a monolithic unit.  Second, system reliability is improved because of the 
modularity that allows any one quadrant to fail without bringing the observatory capabilities below the science requirements.  Finally, 
the composite optical bench has reduced mass, increased stiffness and a reduced coefficient of thermal expansion relative to 
earlier aluminum structural designs.  We describe the individual instruments in the sections below.

\begin{figure}
\includegraphics[width=3.25in]{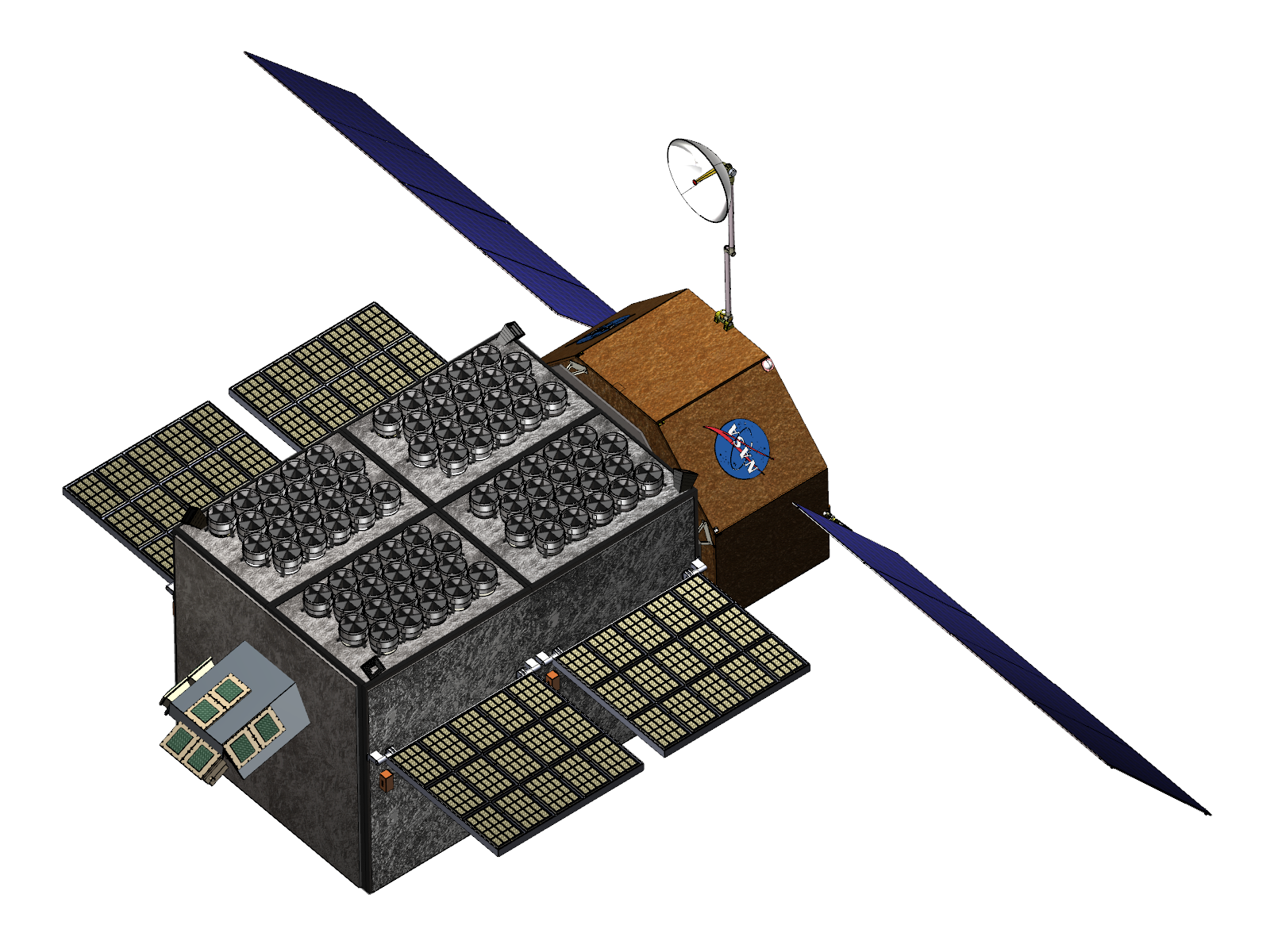}
\includegraphics[width=3.25in]{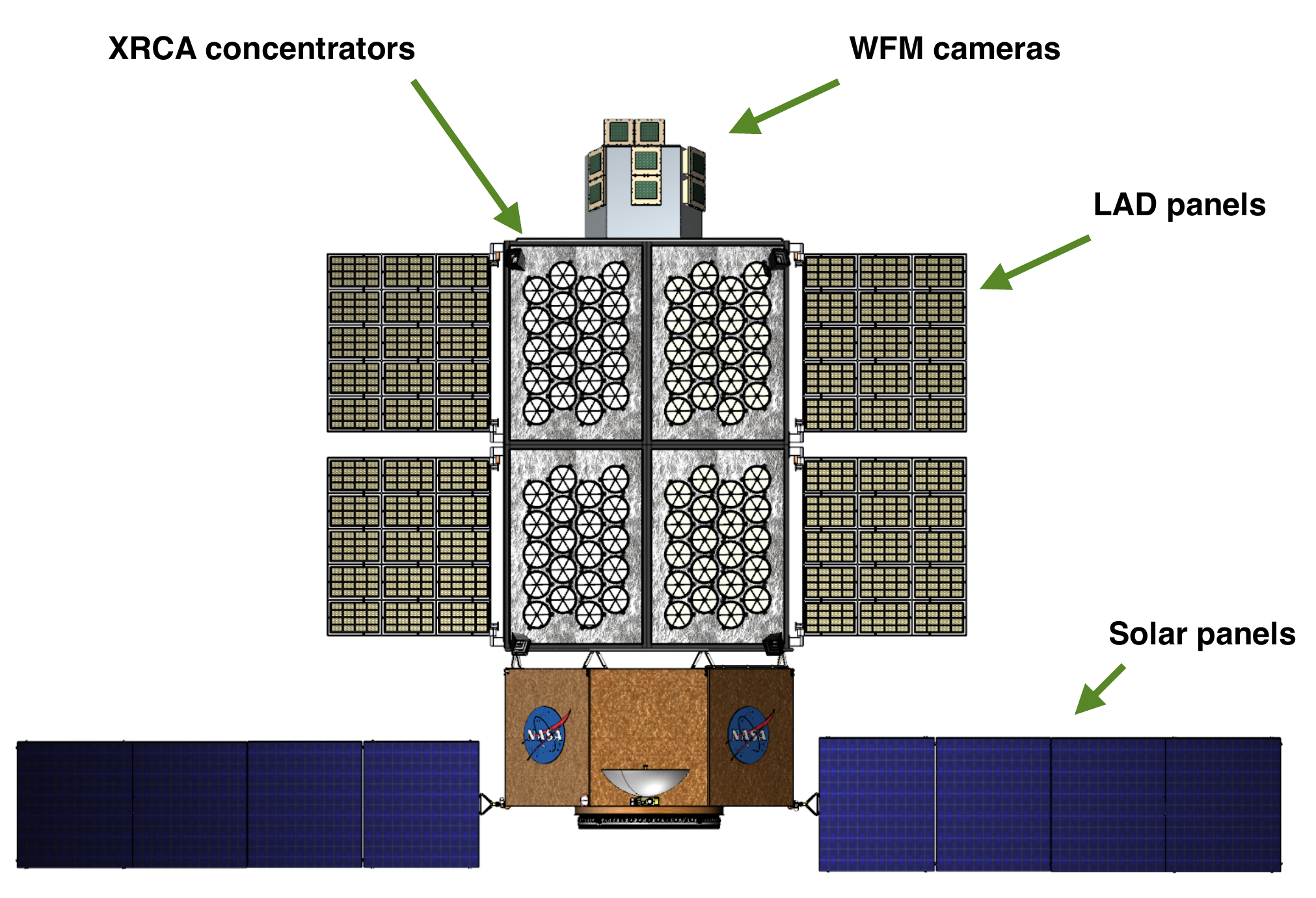}
\caption{Detailed design renderings of the {\em STROBE-X} mission from the NASA/GSFC Instrument Design Laboratory (IDL) and Mission Design Laboratory (MDL).\label{fig:strobex}}
\vspace*{0.1in}
\end{figure}

\subsection{X-ray Concentrator Array (XRCA)}

\begin{figure}
\includegraphics[width=3.0in]{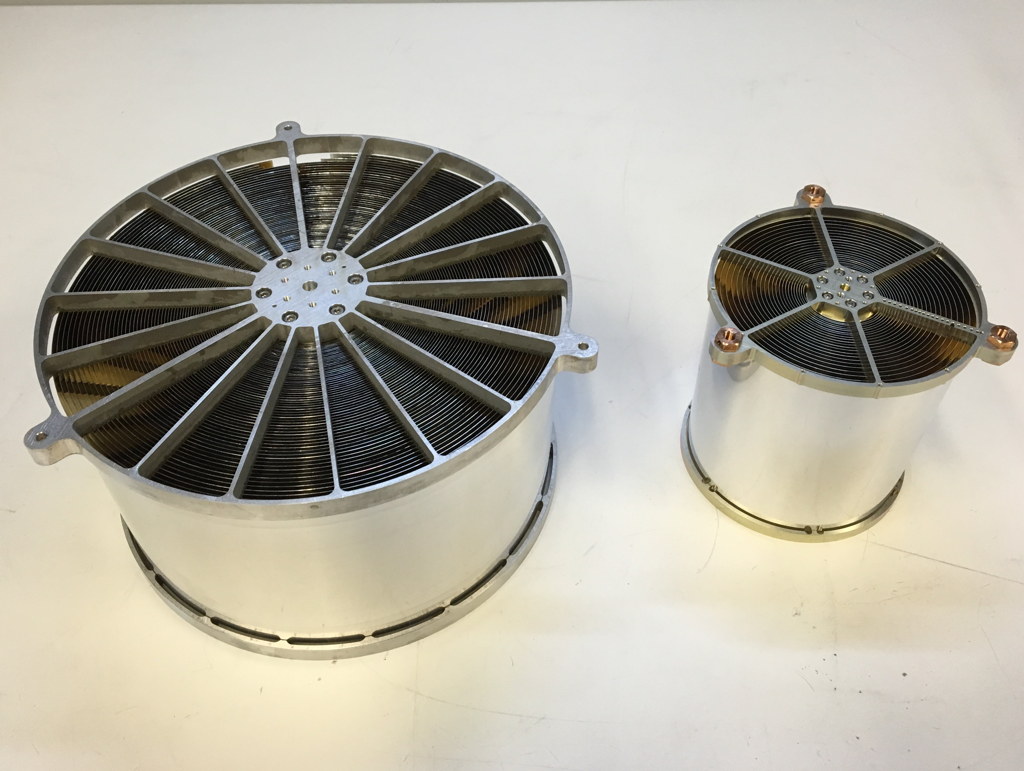}
\hfill\includegraphics[width=3.5in]{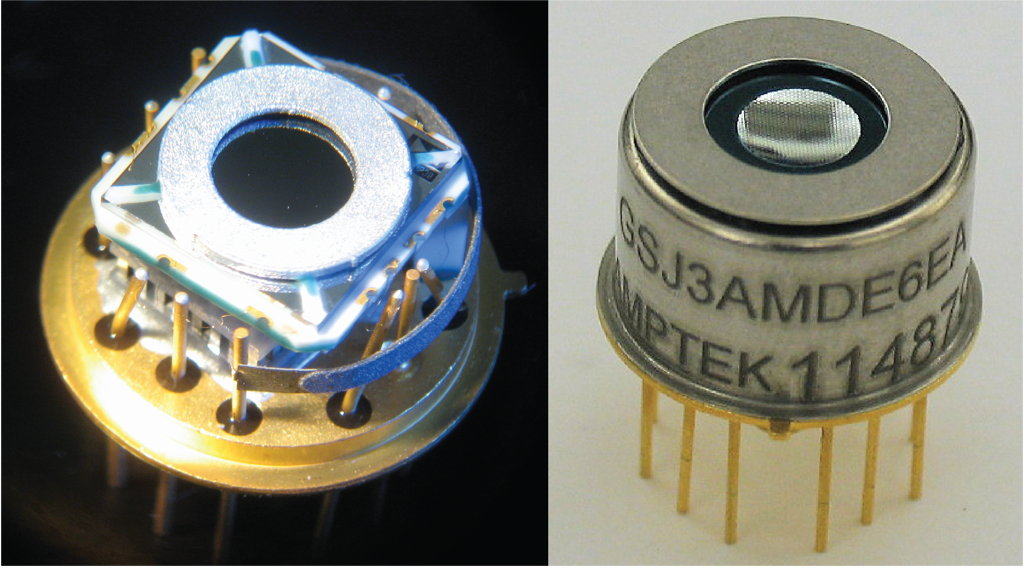}
\caption{Optics and detector technology for {\em STROBE-X}/XRCA. Left: X-ray concentrators from \textit{XACT} (larger) and \textit{NICER} (smaller). \strobex{} would use optics very similar to \textit{XACT}, so no new technology development is required. Right: Si drift detectors from \textit{NICER}, which meet all the requirements for \strobex{}.\label{fig:XRCAopt}}
\end{figure}

\strobex\ covers the soft X-ray (0.2--12 keV) band with the XRCA instrument, a modular
collection of identical X-ray ``concentrator'' (XRC) units that
leverage the successful design and development efforts associated
with GSFC's X-ray Advanced Concepts Testbed (\textit{XACT}) sounding-rocket
payload \cite{doi:10.1117/12.926152} and the \textit{NICER} mission of opportunity
\cite{doi:10.1117/12.2231304,doi:10.1117/12.2234436}. Flight-like
builds of the \textit{XACT} and \textit{NICER} XRCs are pictured in
Fig.~\ref{fig:XRCAopt} (left). 

A concentrator is a high-throughput optic that is optimized for
collecting photons from a point-like (less than $\sim 2$ arcmin in
extent) source over a large geometric area and delivering them onto
small detectors. With reduced detector size, particle interactions
that mimic cosmic X-ray detections are minimized, reducing
background by orders of magnitude. 

The design of the XRCA's concentrators benefits from technological 
heritage and manufacturing experience spanning decades of X-ray
astronomy missions, including \textit{BBXRT}, the \textit{SXS}
sounding-rocket payload, \textit{ASCA}, \textit{Suzaku},
\textit{InFocus}, \textit{Hitomi}, and \textit{NICER}. These optics focus X-rays
using grazing-incidence reflections. The individual optical elements
are nested aluminum-foil shells that are inexpensive to fabricate
and provide high reflectivity in soft X-rays by virtue of a smooth
($\sim 5$\,\AA\ roughness) replicated gold surface. The diameters
and paraboloidal figures of the shells are chosen so that graze
angles do not exceed approximately $2^\circ$.

As implemented on \textit{XACT} and \textit{NICER}, XRCs depart from the
GSFC imaging optics of earlier missions in two ways that enhance
their throughput and point-spread function (PSF) performance, with
no added risk. First, they are formed to have paraboloidal figure
instead of conical approximations, an enhancement that significantly
improves their PSF and vignetting characteristics, with the
attendant benefit of reduced background as detector apertures can be
made smaller. Second, because imaging is not a requirement,
concentrators eliminate the secondary mirrors that are needed for
true imaging optics. 
Thus, 1) they only suffer reflection inefficiencies once, resulting
in enhanced effective area; 2) the number of optical elements
required is half that of an imaging configuration, resulting in
substantial cost and schedule savings; and 3) with no need to align
primary and secondary optics, integration is significantly
simplified.

The \strobex\ XRCA is four quadrants of 20 identical concentrator units each (i.e. 80 units total), which are scaled-up versions of the 56 \textit{NICER} X-ray Timing Instrument
concentrators. Each XRC has a focal length of 3.0 m, with a set of
107 nested foil shells spanning a range of diameters between 3 cm
and 28 cm. The foils are held in place by a spoked-wheel (or
``spider'') structure with mount points that are used to adjust the
XRC alignment in tip and tilt. 

The \strobex\ XRC design retains approximately the same
focal ratio as \textit{NICER}'s optics, so that $\sim 2$ arcmin on-axis focal
spots are again achieved, while the longer \strobex\ focal length
enhances throughput at energies above 2.5 keV.  Detectors will be
masked with apertures corresponding to a 4 arcmin diameter FOV, to
fully capture the PSF while minimizing diffuse sky background
focused into the aperture.

In its baseline configuration, the \strobex\ XRCA adopts \textit{NICER}'s 
silicon-drift detectors (SDDs) and readout 
electronics \cite{doi:10.1117/12.2231718}.  Within these SDDs, a 
radial electric field guides ionization charge clouds from a large (25 
mm$^2$) area to a central low-capacitance readout anode; the resulting 
charge pulse is amplified and shaped to enable measurement of its 
height and the time at which it triggers digitization and further 
processing. Built-in thermoelectric coolers (TECs) maintain each 
detector at $-55^\circ$ C to minimize thermal electron 
noise. The \strobex\ SDDs offer CCD-like energy resolution, 85 to 175 
eV FWHM over the 0.2--10 keV range. They also enable
very precise photon detection time-stamping, well 
under 100 ns RMS. The detectors are thick enough to 
provide $\sim 50$\% quantum efficiency at 15 keV, and are packaged 
with an aluminized thin-film window that offers good transparency to 
photon energies as low as 0.2 keV while maintaining a hermetic 
seal.

The XRCA detector architecture consists of analog electronics 
including a charge-sensitive preamplifier 
integrated within the SDD assemblies,
and power, TEC control, and digital electronics  
that communicate X-ray event data (arrival time, photon energy, 
and event quality) to the observatory's command and data-handling 
system for multiple detectors---e.g., groups of eight---simultaneously. 
The XRCA is capable of 
supporting high photon count rates by virtue of both its modularity and 
the fast readout characteristics of the individual SDD channels: even 
with the seven-fold increase in collecting area of a single \strobex\ XRC 
relative to \textit{NICER}, an SDD module built to \textit{NICER} specifications would 
not be affected by pile-up for incident fluxes below 2 crab, and this 
performance can be improved further with modest changes to the
readout electronics.

\begin{table}
\caption{Parameters for the LAD and XRCA instruments\label{fig:ladxrca}}
\includegraphics[width=6.5in]{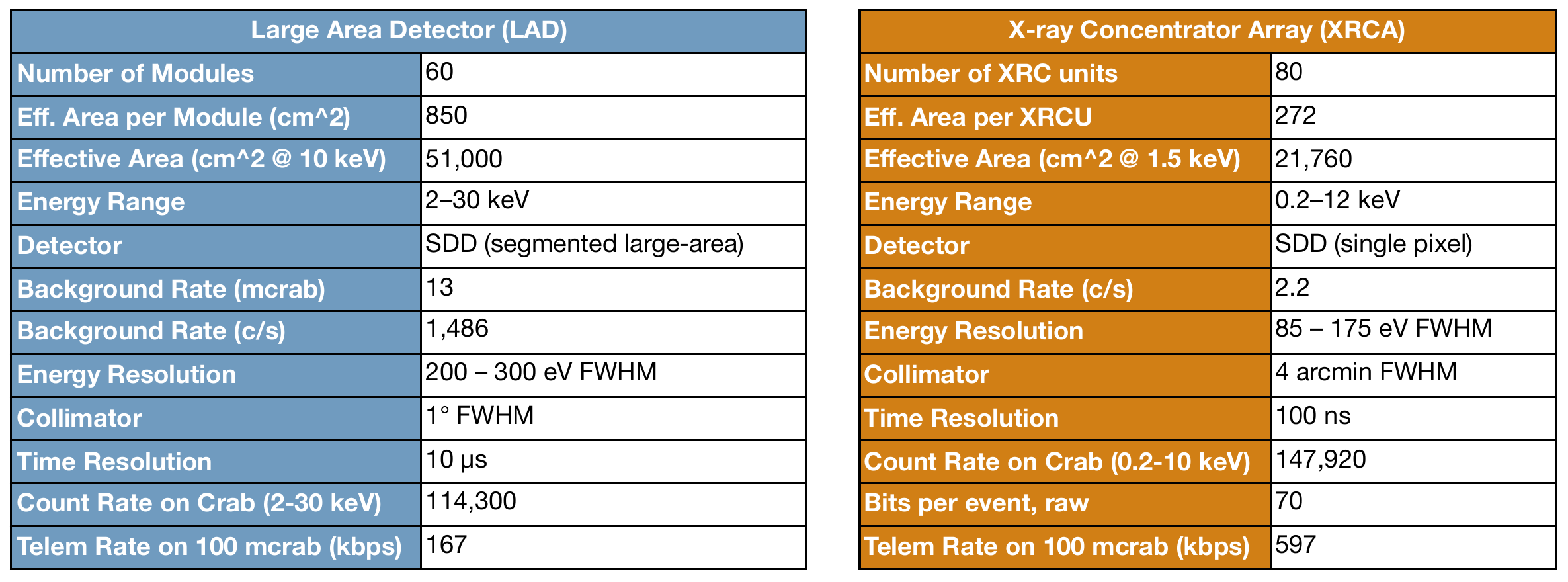}
\end{table}


\subsection{Large Area Detector (LAD)}

The Large Area Detector (LAD) is a large-area, collimated instrument, operating in the 2--30 keV
nominal energy range. The instrument is based on the technologies of the large-area Silicon Drift 
Detectors (SDD) and capillary plate collimators, enabling several square meters to be deployed in space
within reasonable mass, volume and power budgets. The concept and design of the LAD instrument is 
based on the same instrument proposed as part of the scientific payload of the {\em LOFT} mission concept 
\cite{LOFTExpA,Zane_etal_2014}. 

\begin{figure}
\centering
\includegraphics[width=4.0in]{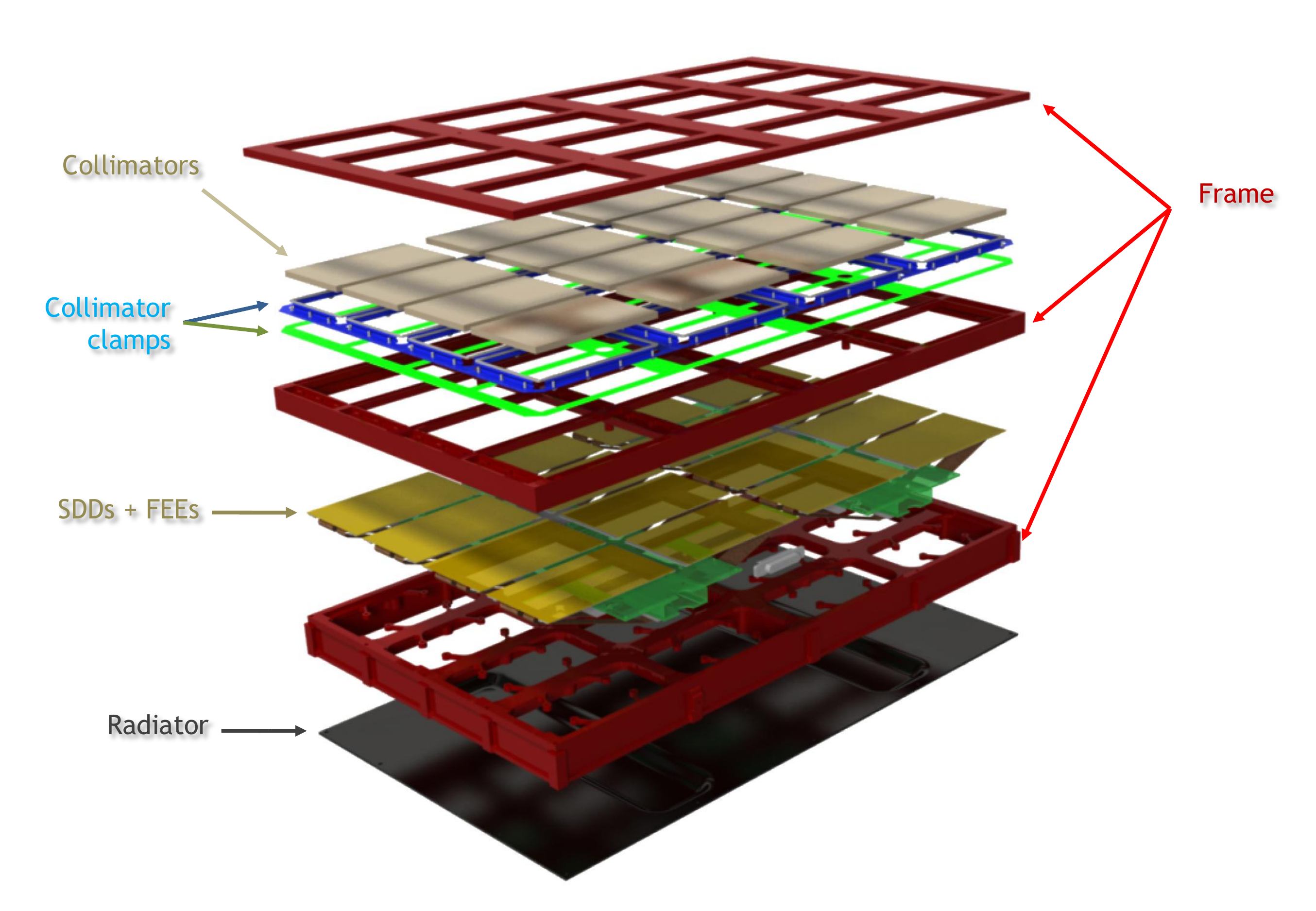}
\caption{An exploded view of the {\em STROBE-X}/LAD module frame design, based on the design for the ESA {\em LOFT} mission concept.
\label{fig:module_with_legend}}
\end{figure}

The LAD's unprecedented collecting area is achieved through a modular and intrinsically highly redundant design. Each LAD Module hosts a set of 4~$\times$~4 detectors with their front-end electronics  and 4~$\times$~4 collimators, supported by two grid-like frames. The back-side of the Module
hosts the Module Back-End Electronics (MBEE). It controls SDDs, FEEs and Power Supply Unit (PSU), 
reads-out the FEE digitized events, generates housekeeping and ratemeters, formats and time-stamps 
each event and transmits it to the Panel Back End Electronics (PBEE). A 300 $\mu$m Pb back-shield 
and a 2 mm Al radiator complete the Module structure, with the tasks of reducing the background events and dissipating heat from Module box, respectively. An exploded view of the LAD Module and its components is shown in Figure \ref{fig:module_with_legend}.

The LAD Modules are organized in 4 large Panels, deployable from each of the {\em STROBE-X} 
quadrants (see {Fig.~\ref{fig:strobex}}). The total effective area is about 5m$^{2}$ at 8 keV.
Each of the LAD Panels hosts 15 (5~$\times$~3) Modules, for a total of 60 Modules or 960 detectors, 
and a PBEE, for a total of 4 PBEEs, in charge of interfacing the 15 Modules to the central 
Instrument Control Unit (ICU). The main parameters of the LAD are listed in Table~\ref{fig:ladxrca}.

\begin{figure}[t!]
\centering\includegraphics[width=4.0in]{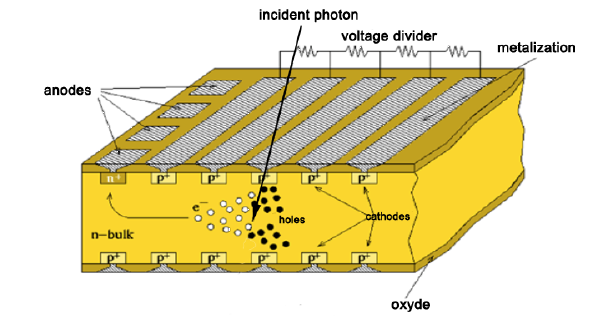}
\caption{Working principle of the Si drift detector used for the {\em STROBE-X}/LAD (see text for details).}

\label{fig:sdd_working_principle}
\end{figure}

The design of such a large instrument is feasible thanks to the detector technology  
of the large-area Silicon Drift Detectors (SDDs, \cite{gatti84}), developed for the ALICE/LHC experiment 
at CERN \cite{Vacchi_etal_1991} and later optimized for the detection of photons to be used on 
{\em LOFT} \cite{2014JInst...9P7014R}, with typical size of 11~$\times$~7 cm$^{2}$ and 450 $\mu$m thickness. 
The key properties of the Si drift detectors  are their capability to read-out a large photon collecting 
area with a small set of low-capacitance (thus low-noise) anodes and their very low mass 
($\sim$1 kg m$^{-2}$). The working principle is shown in Figure \ref{fig:sdd_working_principle}: 
the cloud of electrons generated by the interaction of an X-ray photon is drifted towards the read-out 
anodes, driven by a constant electric field sustained by a progressively decreasing negative voltage 
applied to a series of cathodes, down to the anodes at $\sim$0 V. The diffusion in Si causes the 
electron cloud to expand during the drift. The charge distribution over the collecting anodes depends on the absorption point in the detector. The maximum drift time of $\sim$ 7 $\mu$s, which is the highest detector contribution to the uncertainty in the determination of the arrival time of the photon. Each LAD detector is segmented in two halves, with 2 series of 112 read-out anodes (with 970 $\mu$m pitch) at two edges and the highest voltage along its symmetry axis. 

The high-density anodes of the detector require a read-out system based on ASICs. The requirement on the energy
resolution implies that low-noise and low-power ASICs are needed (17 e$^{-}$ rms noise with 650 $\mu$W\/channel). The read-out of each LAD SDD detector is performed by 8 full-custom 32-channel ASICs inherited from the IDeF-X HD development\cite{2012NIMPA.695..415G}, with A/D conversion carried out by one 16-channel OWB-1 ASIC \cite{2017ITNS...64.1071B}. The dynamic range of the read-out electronics is required to record events with energy up to 80 keV. The events in the nominal energy range (2--30 keV) are transmitted with 60 eV energy binning, while those in the “expanded” energy range, 30-80 keV, are transmitted with reduced energy information (2 keV bins) as they will be used to study only the timing properties of bright/hard events (e.g., gamma-ray bursts, magnetar flares) from outside the field of view, through the collimator structure. 
Despite detecting as many as 120,000 counts per second from the Crab, the segmentation into 960 detectors 
and 215,000 electronics channels means that the rate on the individual channel is very low even for very bright sources, removing any pile-up or dead-time issues.  To maintain good energy resolution throughout the mission lifetime, the detectors need to be cooled to reduce the leakage current.  To keep the energy resolution below 300 eV at end-of-life, the detector temperature must be kept below $-30^{\circ}$C.  Operating at higher temperatures is allowable, but the energy resolution will be degraded. Passive cooling is to be used, given the large size of the instrument.

Taking full advantage of the compact detector design requires a similarly compact collimator design. This is provided by the capillary plate technology.
In the LAD geometry, the capillary plate is a 5~mm thick sheet of lead-glass ($>$40$\%$ Pb mass fraction) with same size as the SDD detector, with round micro-pores 83 $\mu$m in diameter, limiting
the field of view (FoV) to 0.95$^{\circ}$ (full width at half maximum). The open area ratio of the device is 75$\%$. 
The thermal and optical design is then completed by an additional optical filter, composed by a thin (1 $\mu$m thickness) 
Kapton foil coated with $>$100 nm aluminium. This guarantees 10$^{-6}$ filtering on IR$/$Visible$/$UV light, while transmitting $>$90$\%$ at 2 keV and above.\\

\subsection{Wide Field Monitor (WFM)}

\begin{figure}[t!]
\includegraphics[width=2.5in]{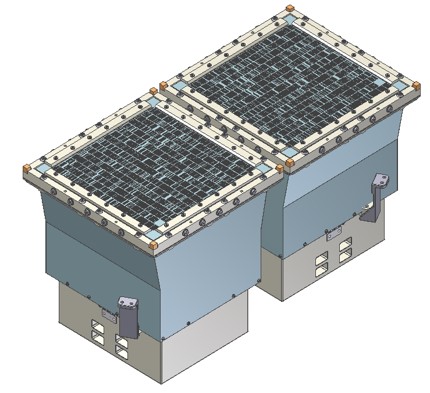}\hfill
\includegraphics[width=4.0in]{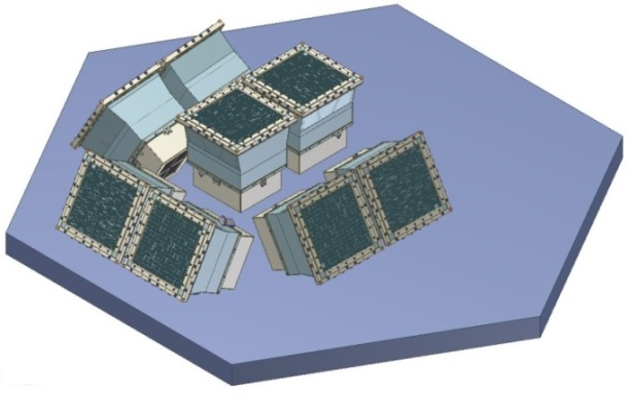}
\caption{The {\em STROBE-X}/WFM instrument. Left: A WFM camera pair. Right: The WFM assembly with four camera pairs.}
\label{fig:WFM_cameras}
\end{figure}

The Wide Field Monitor (WFM) is a coded mask instrument consisting of four pairs of identical cameras, with position sensitive detectors in the (2--50) keV energy range. The same Silicon Drift Detectors (SDDs) of the LAD are used, with a modified geometry to get better spatial resolution. These detectors provide accurate positions in one direction but only coarse positional information in the other one (1.5D). Pairs of two orthogonal cameras are used to obtain precise two-dimensional (2D) source positions (see Fig. \ref{fig:WFM_cameras}, left). The design of the WFM is modular, so that there is no need to put the two cameras of each camera pair together. The concept and design of the WFM is inherited from the {\em LOFT} WFM instrument \cite{LOFTExpA,WFM2014}. 

The effective field of view (FoV) of each camera pair is about 70$^{\circ} \times 70^{\circ}$ ($30^{\circ} \times 30^{\circ}$ fully illuminated, $90^{\circ} \times 90^{\circ}$ at zero response). A set of four camera pairs is foreseen, with three pairs forming an arc covering 180$^{\circ}$  along the the sky band accessible to the LAD and XRCA, and the fourth pair aimed to monitor the anti-Sun direction (see Fig. \ref{fig:WFM_cameras}, right). 

\begin{figure}[t!]
\centering\includegraphics[width=5.5in]{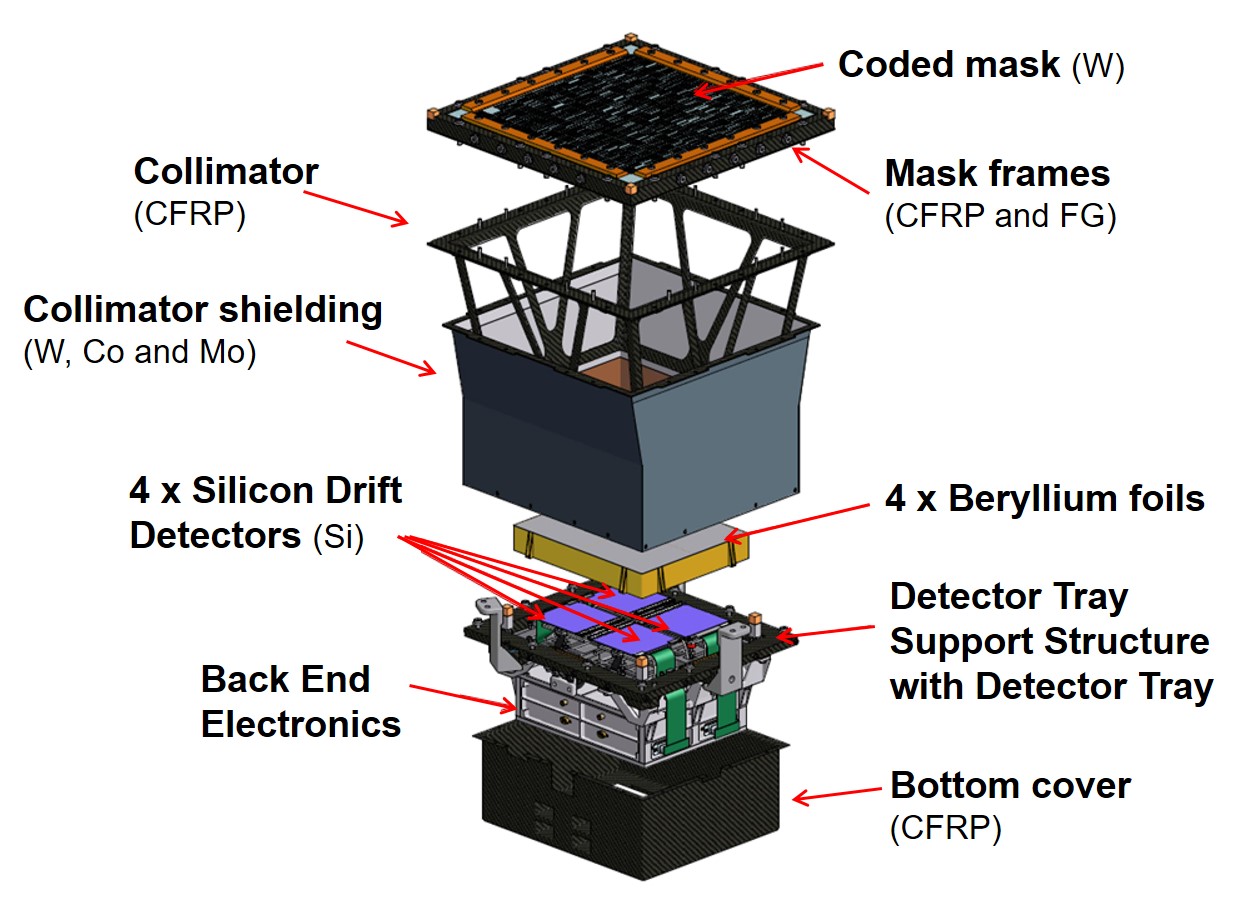}
\caption{An exploded view of a {\em STROBE-X}/WFM camera, indicating all its components.}
\label{fig:WFM_exploded_view}
\end{figure}

Each camera includes a detector tray with four Silicon Drift Detectors, four Front-End Electronics, four Be windows, one Back End Electronics assembly, a Collimator and a Coded Mask with a Thermal Blanket, as shown in Fig. \ref{fig:WFM_exploded_view}. In addition, two Instrument Control Units (ICUs), in cold redundancy, are required. 

The WFM SDDs, ASICs and Front End Electronics are similar to those of the LAD, except that the SDD anode pitch is reduced (145 $\mu$m versus 970 $\mu$m) to improve spatial resolution. There is a higher number of ASICs per SDD: 28$\times$ IDeF-X HD ASICs, with smaller pitch, and 2$\times$ OWB-1 ASICs. Also the PSU (Power Supply Unit) and the Back End Electronics are similar to those for the LAD, but with the BEE providing additional capability to determine photon positions. The ICU controls the eight cameras independently, interfaces with the Power Distribution Unit, and performs on board location of bright transient events in real time.

A beryllium window above the SDDs, 25 $\mu$m thick, is needed to prevent impacts of micro-meteorites and small orbital debris particles (see Fig. \ref{fig:WFM_exploded_view}).

The coded mask of each WFM camera is made of tungsten, with an area of $260\times 260$ mm$^2$ and a thickness of 150 $\mu$m. The mask pattern consists of $1040 \times 16$ open/closed elements, with a mask pitch of 250 $\mu$m $\times$ 14 mm. The dimensions of the open elements are 250 $\mu$m $\times$ 16.4 mm, with 2.4 mm spacing between the elements in the coarse resolution direction for mechanical reasons. The nominal open fraction of the mask is 25$\%$. The detector-mask distance is 202.9 mm. The corresponding angular resolution (FWHM) for the on-axis viewing direction is equal to the ratio of the mask pitch to the detector to mask distance: 4.24 arcmin in the high-resolution direction and 4.6 degrees in the coarse resolution direction. 
The mask must be flat, or at least maintain its shape, to $\pm 50$ $\mu$m over its entire surface across its full operational temperature range. This is the main reason why a Sun shield is required for the WFM. The design of the mask frame also helps to fulfill this requirement.

The collimator supports the mask frame and protects the detector zone from  X-rays coming from outside. It is made of a 3 mm thick open Carbon Fiber Reinforced Plastic (CFRP) structure, covered by a shield with the following structure: a 1mm thick CFRP layer outside covered by a 150 $\mu$m Tungsten foil, and inside covered by Cu and Mo 50 $\mu$m thick foils. Cu and Mo are included for in-flight calibration purposes (see Fig. \ref{fig:WFM_exploded_view}). 

The BEE box is located at the bottom of the camera. It includes the BEE board and the power supply unit. It is protected by a CFRP cover. The ICU includes the WFM Data Handling Unit, with mass memory and the Power Distribution Unit; there is a main and a redundant unit, in separate boxes.  

\begin{table}
\caption{Parameters for the WFM instrument and the overall {\em STROBE-X} Mission \label{fig:wfmstrobex}}
\includegraphics[width=6.5in]{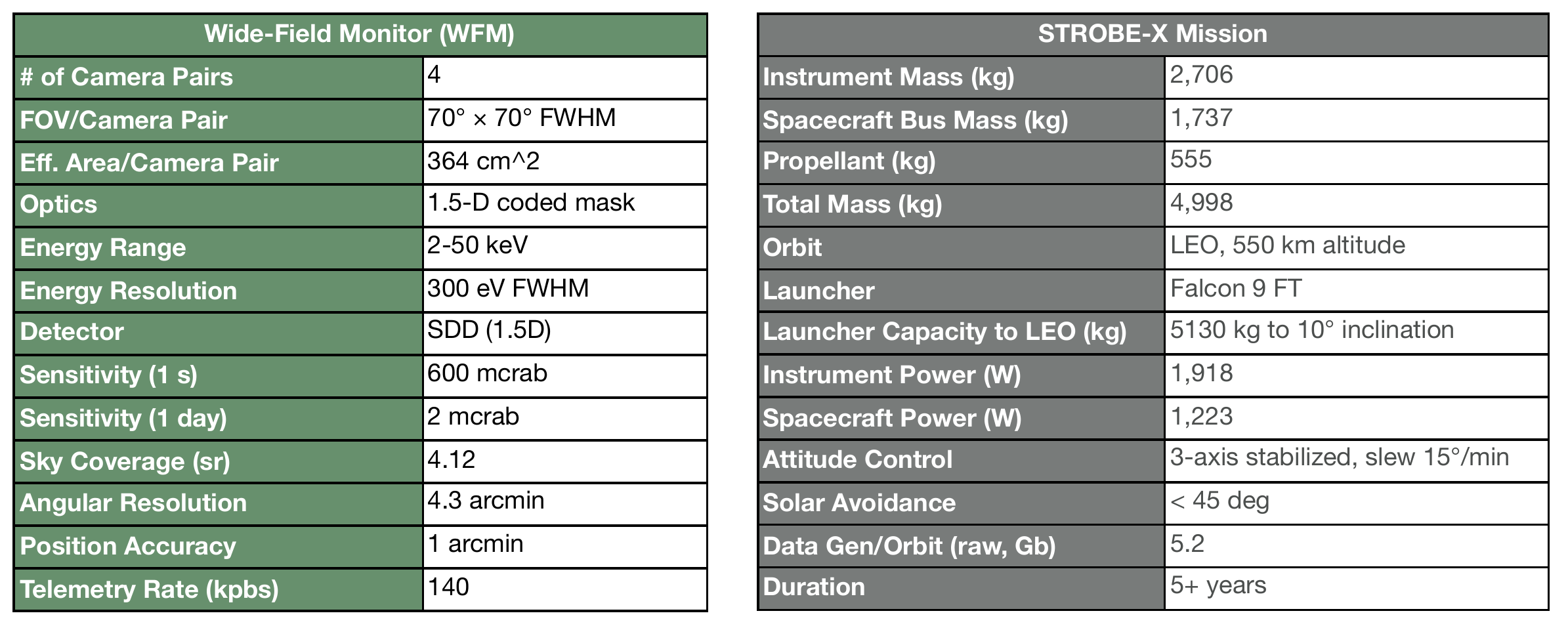}
\end{table}


\section{MISSION}

The overall mission concept is that of an agile X-ray observatory in low-Earth orbit, similar to previous missions like {\em RXTE} and 
{\em Swift}. A study in 2018 April at the NASA/GSFC Mission Design Lab (MDL) developed the spacecraft bus design and other aspects of the mission, as described below.

\subsection{Launch and Orbit}
The LAD detectors are sensitive to non-ionizing energy losses from radiation exposure (which causes increased leakage current), so minimizing the time spent in the 
South Atlantic Anomaly (SAA) by going to as low an orbital inclination as possible is desirable. 
We evaluated {\em STROBE-X} using the performance parameters provided by SpaceX via NASA Launch Services for a Falcon 9 launch vehicle, 
assuming an expendable first stage. The mass that can be put into a 600 km circular orbit is a very strong function of the
desired inclination, with a capacity of 5130 kg to 10$^\circ$ inclination and 7730 kg to 15$^\circ$ inclination. We thus 
plan for a 550 km altitude circular orbit at an inclination of 10$^\circ$.  If additional mass margin is needed, a small 
inclination increase can easily allow the launcher to accommodate that. And, if an equatorial launch site becomes available
from SpaceX or as a European contribution, then {\em STROBE-X} could avoid the SAA entirely, which would increase efficiency and
reduce the cooling requirements on the LAD.

\subsection{Propulsion}

Avoiding a propulsion system altogether would be desirable from a cost and complexity perspective. 
However, there are three reasons that one could be necessary: (1) if the casualty probability from an 
uncontrolled reentry exceeds $10^{-4}$, (2) if reboosting is required to achieve the goal of 10-year orbital lifetime, 
or (3) if a maneuver capability is needed to avoid orbital debris collisions. While none of these are formally required,
we have taken the conservative approach and included a propulsion system capable of all three.

\subsection{Pointing and Attitude Control}

{\em STROBE-X} must be able to slew rapidly over the full sky outside of the 45$^\circ$ Sun avoidance region in order to
follow transients, make monitoring observations, and respond rapidly to targets of opportunity. The minimum required
slew rate is 5$^\circ$/minute, with a goal of 15$^\circ$/minute. While the minimum rate could be achieved with conventional
reaction wheels, this would be inefficient for short observations and delay getting to fast transients. We thus chose
to use Honeywell M50 control moment 
gyroscopes\footnote{\url{https://aerocontent.honeywell.com/aero/common/documents/myaerospacecatalog-documents/M50_Control_Moment_Gyroscope.pdf}} 
(CMGs) for attitude control. These are somewhat more expensive than 
reaction wheels but allow us to reach our 15$^\circ$/minute goal, and are at high TRL. 
CMGs have somewhat larger jitter than reaction wheels, which can be an issue for high-resolution imaging instruments, but STROBE-X's instruments only require arcminute-scale stability.

Attitude knowledge is provided by star trackers and coarse Sun sensors and momentum unloading is accomplished with magnetic torquers.

Most maneuvers are planned on the ground and uploaded as time tagged commands, but the spacecraft also has the ability
to perform an autonomous repoint based on a message from the WFM processor, which can be programmed to trigger on specific
transients, like GRBs and superbursts.


\subsection{Telemetry and Data Rates}

We use TDRSS Ka band downlink via a high gain antenna to achieve 300 Mbps. This enables downlink of an estimated 540 Gb/day average.  A key driver for telemetry capacity is the ability to observe bright sources with full energy and time resolution. Table 1 shows telemetry rates for the LAD and XRCA, assuming they are observing a typical 100 mcrab source, resulting in an easily accommodated telemetry rate of of 5.2 Gb/orbit. To estimate the telemetry capacity needed to accommodate full resolution observations of bright sources and to allow more extended-time observations of bright sources to be downlinked over several orbits, we assumed that the two {\em STROBE-X} pointed instruments are observing a 1 mcrab source 25\% of the time, a 500 mcrab source 50\% of the time, a 5 Crab source 5\% of the time, and background 20\% of the time, with 15 minutes per orbit in SAA for an entire day, resulting in 540 Gb/day. This high telemetry capacity enables {\em STROBE-X} to downlink all events all the time, including for very bright sources, and all events for the WFM.  This capability is an important enhancement relative to the {\em LOFT} mission concept.

TDRSS S-band Multiple Access through a pair of omnidirectional antennas allows broadcasting of burst and transient 
alerts to the ground in less than 10 seconds as well as rapid commanding from the ground 
in response to a TOO request. The burst and transient alerts will be rapidly followed by localizations and quick look light curves, similar to {\em Swift} and {\em Fermi}/GBM. These S-band antennas can also communicate with ground stations for contingencies and 
launch and early orbit operations.

\subsection{Cost and Schedule}

To be considered as a candidate Probe class mission, NASA requires that the total lifecycle mission cost estimate (Phases A--F)
be less than \$1B in FY 2018 dollars. NASA has provided guidance that: (1) \$150M should be held for launch costs, 
(2) unencumbered cost reserves should be 25\% of Phases A/B/C/D costs, 
(3) cost assumptions should be for unmodified Class B missions, (4) assume a Phase A start date of 2023 October 1.
The IDL and MDL studies provided detailed cost estimates for the instruments, spacecraft bus, ground systems, integration and 
test, and downlink costs, using a combination of parametric and grass roots costing. To that, we have added standard
percentage ``wraps'' for WBS items like Project Management, Systems Engineering, Safety and Mission Assurance, Science, and 
Education and Public Outreach. For mission operations, we found that the wrap was a significant underestimate compared to NASA
guidance and historical precedent from missions like \textit{Fermi}, so we increased it to \$15M/year for the prime mission. This process
resulted in a total cost estimate comfortably below the Probe class cost cap, with over 10\% margin (above and beyond the mandated 25\% reserves), giving us very high confidence that this mission can be executed as a Probe.  We note that this cost estimate is very conservative in that it assumes all costs are borne by NASA. In reality, there would surely be a significant contribution from Europe that would reduce the cost to NASA.

For the mission schedule, we followed the NASA guidance for Probe class missions being proposed to the 2020 Decadal Survey and
used a Phase A start date of 2023 October 1. Based on this, we constructed a detailed construction, integration and test 
plan, including 8 months of funded schedule reserve, giving an estimated launch date of 2031 January 1.


\acknowledgments 
 
The {\em STROBE-X} mission concept study is funded by NASA. 
The Italian authors acknowledge support from ASI, under agreement 
ASI-INAF n.2017-14-H.O, INAF and INFN. The Spanish authors acknowledge support from MINECO grant ESP2017-82674-R and FEDER funds. ALW acknowledges support from ERC Starting Grant 639217.

We gratefully acknowledge the superb engineering teams at the NASA/GSFC Instrument Design Lab and Mission Design Lab for their important contributions to the {\em STROBE-X} technical concept. We thank Lorenzo Amati and Giulia Stratta for useful discussions.

\bibliographystyle{spiebib} 
\bibliography{report} 


\end{document}